\documentclass[]{mn2e}

\usepackage{epsfig,mathptm}

\def\be{\begin{equation}} 
\def\ee{\end{equation}} 
\def\ba{\begin{eqnarray}} 
\def\ea{\end{eqnarray}}

\def\reff@jnl#1{{\rm#1\/}}
\def\aj{\reff@jnl{AJ}}                  % Astronomical Journal
\def\araa{\reff@jnl{ARA\&A}}            % Annual Review of Astron and Astrophys
\def\apj{\reff@jnl{ApJ}}                        % Astrophysical Journal
\def\apjl{\reff@jnl{ApJ}}               % Astrophysical Journal, Letters
\def\apjs{\reff@jnl{ApJS}}              % Astrophysical Journal, Supplement
\def\ao{\reff@jnl{Appl.Optics}}         % Applied Optics
\def\apss{\reff@jnl{Ap\&SS}}            % Astrophysics and Space Science
\def\aap{\reff@jnl{A\&A}}               % Astronomy and Astrophysics
\def\aapr{\reff@jnl{A\&A~Rev.}}         % Astronomy and Astrophysics Reviews
\def\aaps{\reff@jnl{A\&AS}}             
\def\azh{\reff@jnl{AZh}}                        % Astronomicheskii Zhurnal
\def\baas{\reff@jnl{BAAS}}              % Bulletin of the AAS
\def\jrasc{\reff@jnl{JRASC}}            % Journal of the RAS of Canada
\def\memras{\reff@jnl{MmRAS}}           % Memoirs of the RAS
\def\mnras{\reff@jnl{MNRAS}}            % Monthly Notices of the RAS
\def\pra{\reff@jnl{Phys.Rev.A}}         % Physical Review A: General Physics
\def\prb{\reff@jnl{Phys.Rev.B}}         % Physical Review B: Solid State
\def\prc{\reff@jnl{Phys.Rev.C}}         % Physical Review C
\def\prd{\reff@jnl{Phys.Rev.D}}         % Physical Review D
\def\prl{\reff@jnl{Phys.Rev.Lett}}      % Physical Review Letters
\def\pasp{\reff@jnl{PASP}}              % Publications of the ASP
\def\pasj{\reff@jnl{PASJ}}              % Publications of the ASJ
\def\qjras{\reff@jnl{QJRAS}}            % Quarterly Journal of the RAS
\def\skytel{\reff@jnl{S\&T}}            % Sky and Telescope
\def\solphys{\reff@jnl{Solar~Phys.}}    % Solar Physics
\def\sovast{\reff@jnl{Soviet~Ast.}}     % Soviet Astronomy
\def\ssr{\reff@jnl{Space~Sci.Rev.}}     % Space Science Reviews
\def\zap{\reff@jnl{ZAp}}                        % Zeitschrift fuer Astrophysik
\def\nat{\reff@jnl{Nature}}             % Nature 

\title[Submillimetre observations of three $z>6$ quasars]
{Submillimetre observations of {$\bmath{z>6}$} quasars}
\author[Robson et al.]
{Ian Robson$^1$, Robert S. Priddey$^{2,3}$, 
Kate G. Isaak$^{4}$, Richard G. McMahon$^5$\\
$^1$ UK Astronomy Technology Centre, Blackford Hill, 
Edinburgh EH9 3HJ, UK\\
$^2$ Imperial College, Blackett Laboratory, Prince Consort Road, London
SW7 2BZ, UK\\
$^3$ Department of Physics, Astronomy \& Mathematics, University
of Hertfordshire, College Lane, Hatfield AL10 9AB, UK\\
$^4$ School of Physics \& Astronomy, University of Wales - Cardiff ,
 Cardiff, CF24 3YB, UK\\
$^5$ Institute of Astronomy, Madingley Road, Cambridge, CB3 0HA, UK}
\pagerange{\pageref{firstpage}--\pageref{lastpage}}
\pubyear{2004}

\date{Submitted 15th February, 2004; Accepted 15th April, 2004}
\begin{document}
\maketitle
\label{firstpage}

\begin{abstract}
We report on submillimetre (submm) observations of three high redshift
quasars ($z>$6) made using the SCUBA camera on the James Clerk Maxwell
Telescope (JCMT). Only one of the sample was detected ($>10\sigma$
significance) at 850$\mu m$ -- SDSS J1148$+$5251 ($z=6.43$).  It was also
detected at 450$\mu m$ ($>3\sigma$ significance), one of the few
quasars at $z>4$ for which this has been the case.  In combination
with existing millimetric data, the 850$\mu$m and 450$\mu$m detections
allow us to place limits on the temperature of the submm-emitting dust. The
dust temperature is of no trivial importance given the high redshift
of the source, since a cold temperature would signify a large mass of
dust to be synthesized in the little time available (as an extreme
upper limit in only 0.9Gyr since $z=\infty$).  We find, however, that
the combined millimetre and submm data for the source cannot simply be
characterised using the single-temperature greybody fit that has been
used at lower redshifts. We discuss the results of the observing and
modelling, and speculate as to the origin of the deviations.
\end{abstract}


\section{Introduction}
A number of independent lines of investigation over the last 10 years 
have placed submillimetre (submm)  observations of high-redshift quasars 
into the spotlight. Observations have unveiled a population of
extremely luminous submm sources lying at high redshift,
believed to be the dust-obscured, star-forming ancestors of  
massive elliptical galaxies. Contemporaneously, it was realised that a tight
correlation exists between the stellar velocity dispersion of galactic
spheroids, and the mass of their central, supermassive black hole 
(Gebhardt et al. 2000). Taken together, these indicate that luminous 
active galactic nuclei (AGN) at high redshift--- the build-up phase of 
a supermassive black hole --- are prime sites at which to search for the 
dust-enshrouded star-burst phase through which, according to the new 
galaxy-formation paradigm, their massive spheroids necessarily must pass.

The high, sustained luminosity of quasars across the 
electro-magnetic spectrum, allows them to be studied over a wide range 
of both redshift and observing wavelength. Early observations of 
z$\sim$4.5 quasars by McMahon et al. (1994), and 
Isaak et al. (1994) established that some high redshift quasars were prodigious far infrared emitters with 
$\rm L_{fir}\sim10^{13-14} L_{\odot}$ and estimated masses of cool dust of 
$\rm\sim 10^{8-9} M_{\odot}$.

The discovery of 
quasars at $z>6$ (Fan et al. 2003) now makes it possible to 
compile homogeneous, well-defined samples over a significant span 
of the lifetime of the cosmos, from recent times to the threshold 
of reionization. Follow-up is simplified by the accurately-known 
optical positions and the spectroscopic redshifts of the host 
galaxies, which can readily be determined to the precision required 
to pinpoint emission lines from molecular gas--- a key indicator of the 
conditions required for star formation. 

Recent SCUBA studies of the submm emission from high redshift ($z>4$), 
radio-quiet quasars have been reported by 
McMahon et al. (1999), Isaak et al. (2002) and
Priddey et al. (2003b),
along with a sample at lower 
redshift (z$\sim$2) by Priddey et al. (2003a).
A considerable fraction of the targets have been shown to be luminous
submm sources. Interestingly, this fraction appears to have no
significant dependence upon redshift. Similar conclusions have been 
drawn from observations at millimetre (mm) wavelengths 
(eg. Omont et al. (2001), Carilli et al. (2001)).

\section{Observations}
\subsection{The Sample}
Our source-list comprised three of the $z\geq6$ quasars identified by the 
Sloan Digital Sky Survey team, and reported in 
Fan et al. (2003). Source parameters are given in Table~1. Observations
of the other two quasars known (as of January, 2003) to be at $z>6$, 
SDSS J1030$+$0534 ($z=6.28$) and SDSS J1306+0356 ($z=5.99$), 
have been reported in Priddey et al. (2003b).

\subsection{Observations and Data Analysis}

Sources were observed using SCUBA (Holland et al. 1999)
on the JCMT \footnote{The James Clerk Maxwell
Telescope is operated by the Joint Astronomy Centre in Hilo, on behalf of
the parent organisations of the Particle Physics and Astronomy Research 
Council in the UK, the National Research Council of Canada and the 
Netherlands Organisation for Scientific Research}
on the nights of 2003 January 31 and February 01 (UT).
Simultaneous observations were made at 850 and 450$\mu $m using 
photometry mode (placing the source on the central bolometers H7, C14 
of the two arrays respectively) with a 60 arcsec azimuth chop at 
7.8Hz. Data were taken in groups of 40 samples, with observations 
repeated until $S_{850\mu m}\leq 2$mJy was achieved.  
Telescope pointing was checked hourly, and found to be better 
than 3 arcsec. Mars and Uranus, along with IRC$+$10216, were used 
as primary and secondary calibrators respectively. The derived, mean, 
flux conversion factors were 213$\pm 5$Jy/V (850$\mu m$) 
and $355\pm 15$Jy/V (450$\mu m$), with a variation of 
$<10\%$ seen over the course of the observing period. The sky 
opacity was measured using sky-dips and the recently commissioned 
water vapour radiometer that measures the direct line-of-sight atmospheric 
extinction. 

Observing conditions on both nights were moderately good, with $ 0.1 <
\tau^{zen}_{\rm 850} < 0.23$ and $ 0.27 < \tau^{zen}_{\rm 850} < 0.32$
(zenith atmospheric transmission at $850\mu m$ between 79 -- 90\% and
72 -- 76\% ) respectively.  Data were reduced using both the
semi-automated {\rm ORAC-DR} pipeline data reduction package
(Jenness \& Economou 1999)
and a custom reduction procedure based on a
series of routines taken from the {\rm SURF} reduction package
(Jenness \& Lightfoot 1998a,b).
In each case, the
final flux represents the weighted (by the individual rms) average
taken of all data for a particular source/filter combination.

The initial analysis of the SDSS J1148$+$5251 dataset revealed 
considerable variation between both the flux and rms of the individual
40-sample data groups. One explanation of this discrepancy was revealed
upon the subsequent publication of a MAMBO-2 image of 
SDSS J1148$+$5251 by Bertoldi et al. (2003).
It seemed feasible that our fixed, azimuth chop had placed the
off-source position over a second, millimetre-bright source in the
quasar field during the latter stages of our observation.
In order to eliminate this possibility, we therefore obtained subsequent 
SCUBA photometry on SDSS J1148$+$5251, in UK service time
during the nights of 2003 July 9th, 10th and 12th.
This time, a specified chop throw, and position angle fixed relative to 
RA--Dec, were chosen to avoid the potential contaminant sources.
The atmospheric extinction during the observing period 
was low (with $0.14 < \tau^{zenith}_{\rm 850\mu m} < 0.19$) 
and the sky stable. 
Guided by these new, more reliable and consistent observations, 
we reanalysed the initial datasets, testing alternative explanations
for their disagreement.

\section{Results and Discussion}
The measured flux densities of the three sources are tabulated in Table~1, 
along with the observational parameters of the three sources. 
For comparison, the mm fluxes, taken from Bertoldi et al. (2003), 
have also been included. In all cases the numbers in 
brackets are the $1\sigma$ rms values.

\begin{figure}
\vspace{40mm}
\begin{center}
\includegraphics{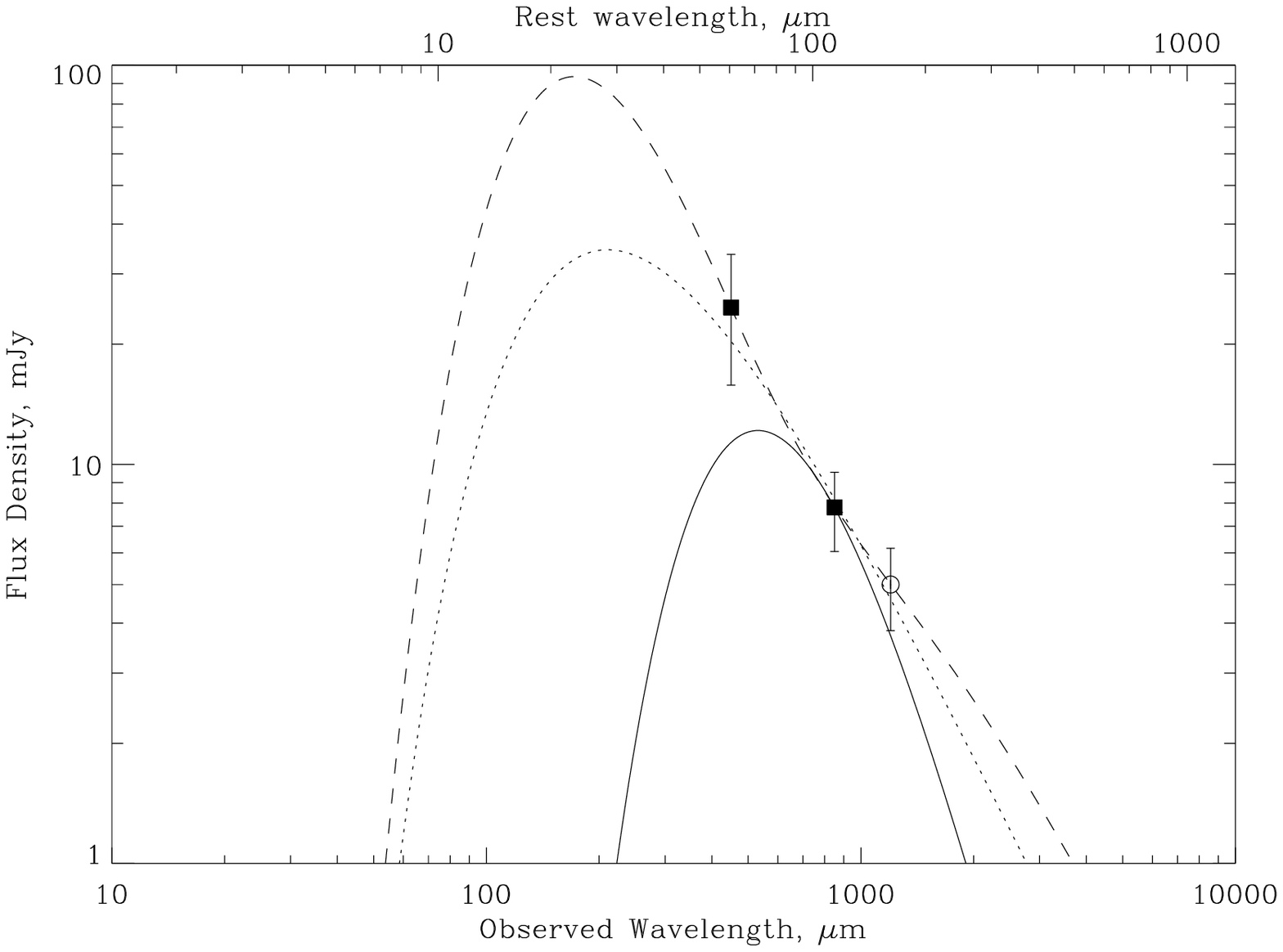}
\vspace{40mm}
\includegraphics{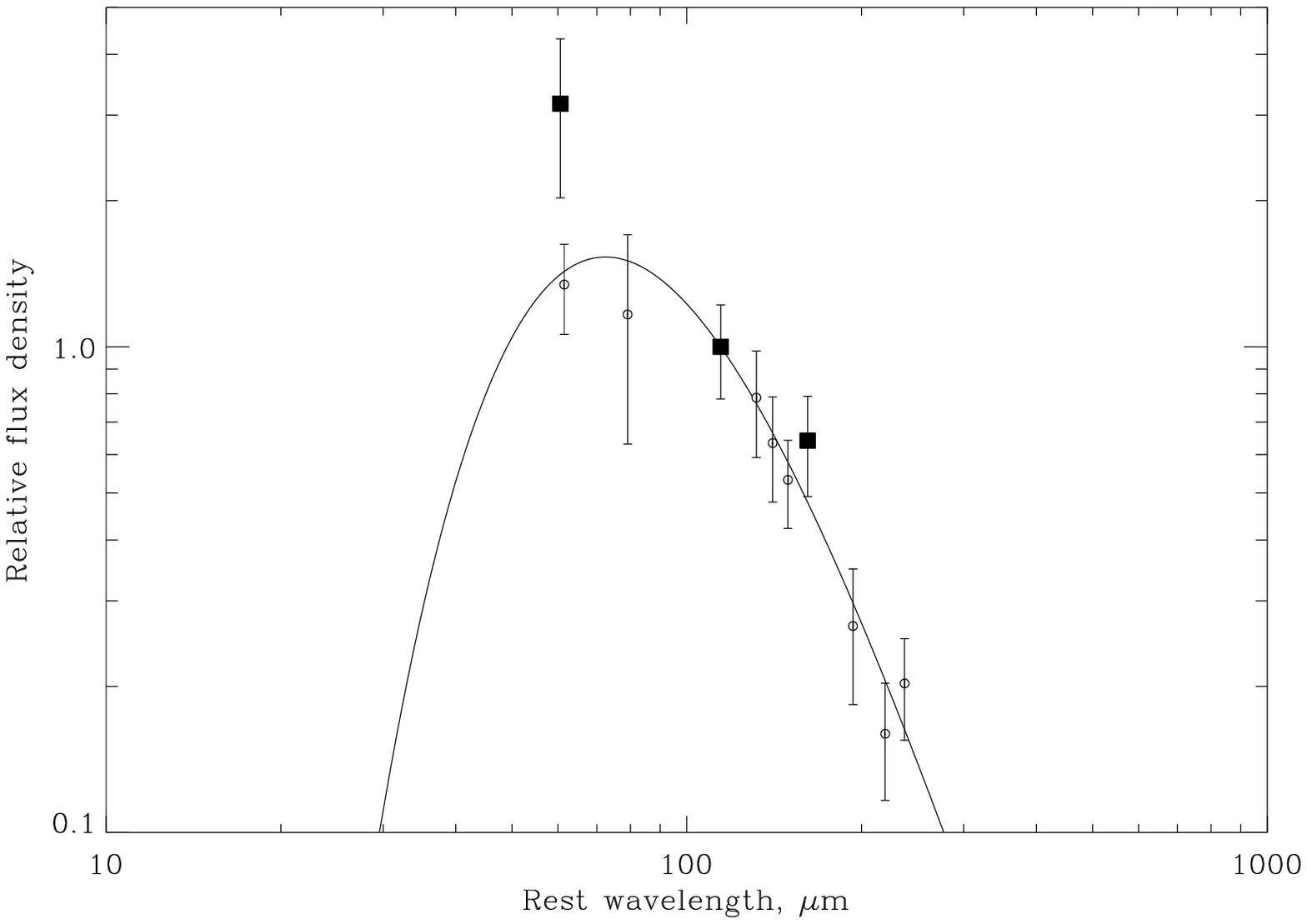}

\vspace{35mm}
\end{center}

\caption{Top panel: 
Observed submm and mm fluxes for SDSS J1148$+$5251. 
Note that for each point, the plotted error bars are the square root of the sum of the squares of the radiometric rms and systematic calibration errors. 
Superposed over the points are three model SEDs: the solid locus traces 
the isothermal dust model of $T$=40K, $\beta=2.0$ 
derived by Priddey\& McMahon (2001) from a sample
of mm and submm observations of $z>4$ quasars, and fit here
through the 850$\mu m$ flux; the dotted 
line traces the SED of a single-temperature dust model with $T_D=180K$, 
$\beta=0$; the dashed line traces a two-component fit 
using ($T_D, \beta$) of (140, 1.5) and (40, 0). The two-component model 
is included for illustrative purposes only, as the number of independent 
data points is smaller than the number of independent constraints on 
the model. \newline
Bottom panel: A plot of the Priddey \& McMahon (2001) SED superposed onto 
the observed mm and submm data fluxes for SDSS J1148$+$5251
(filled squares) and BR B1202$-$0725 (open circles), redshifted to the quasar 
rest frame. The observed fluxes have been normalised to the 
850$\mu m$ values appropriate to the individual sources.} 
\end{figure}

\begin{table*}

\caption{Summary of the source parameters of the SCUBA photometry 
observations}
\begin{tabular}{lccccccccc}
\hline
Source & Redshift  & $M_B$ & RA     &Dec     & $S_{1.2mm}$ 
& Number of   & $S_{850\mu m}$ & $S_{450\mu m}$ \\
       &           &       & (J2000)&(J2000) & (mJy)         & integrations& (mJy)          & (mJy)\\
 (1)   &  (2)      & (3)   & (4)    & (5)    &  (6)        & (7)            & 
 (8)   &  (9)             &  \\
\hline

SDSS J1048$+$4637 & 6.23 & $-$28.15  & 10 48 45.05 & $+$46 37 18.3 & 3.0(0.4) & 120 & 2.3(2.2) & 7.6(11.7)\\

SDSS J1148$+$5251 & 6.43 & $-$28.42  & 11 48 16.64 & $+$52 51 50.3 & 5.0(0.6) & 600 & 7.8(0.7) & 24.7(7.4) \\

SDSS J1630$+$4012 & 6.05 & $-$26.71  & 16 03 33.90 & $+$40 12 09.6 & $3\sigma<1.8$ & 280& 2.7(1.9) & 15.4(9.6) \\

\hline
\end{tabular}
\begin{flushleft}
Positions and optical magnitudes have been taken from Fan et al. (2003);
1.2mm MAMBO-2 data have been taken from Bertoldi et al. (2003)\newline
\end{flushleft}
\end{table*}

\begin{table*}
\caption{Summary of quantities derived from submm and optical photometry.
See text for an explanation of the quantities and notes on their derivation.}

\begin{tabular}{lcccccccccc}
\hline

Source             & $z$   &  $M_B$   & $S_{\rm 850\mu m}$/$S_{1.2mm}$  & $S_{\rm 450\mu m}$/$S_{\rm 850\mu m}$ & $t(\inf) - t(z)$ &  $M_d$ &   $M_{\star}(\rm min)$ & $L_{FIR}$ & $M_{bh}$ & $\dot{M}_{acc}$\\

                   &     &        &                     &                     &   (Gyr)          &  ($10^{8}$M$_{\odot}$)& (M$_{\odot}$/yr) &  (10$^{13}$L$_{\odot}$)

         & 
($10^9$M$_{\odot}$) & (M$_{\odot}$/yr)\\
(1) & (2) & (3) & (4) & (5) & (6) & (7) & (8) & (9) & (10) & (11) \\
\hline 

SDSS J1048$+$4637 & 6.23 & $-$28.15 & $<3.4$                  & $-$& 0.94   & $<4.6$ & $<50$     & $<0.9$ & 6.0 & 130  \\ 

SDSS J1148$+$5251 & 6.43 & $-$28.42 & $1.6^{1.9}_{1.3}$ & $3.2^{4.5}_{2.0}$&0.90 & 5.3 & 60 & 1.1 & 7.7 & 170\\

SDSS J1630$+$4012 & 6.05 & $-$26.71 & $-$   & $-$ & 0.97 & $<4.5$ & $<45$ &$<0.9$ & 1.6 & 35\\

Mean quasar       & 6.24 & $-$27.76   & 2.13                 & 1.65               & 0.94 & $-$ & $-$ & $-$ & 4.2 & 90\\ 

\hline
\end{tabular}

\begin{flushleft}

\end{flushleft}
\end{table*}
Tabulated in Table~2 are the derived 850$\mu m$/1.2mm and 450/850$\mu m$ flux ratios for the sample where sufficient data exist. The superscripts and subscripts in columns 4 and 5 give the $1\sigma$ upper and lower values of the flux ratios. Also listed are the equivalent numbers for the fiducial single-temperature 
SED model of a quasar at the mean redshift of the $z>6$ sample, 
$\overline{z}=6.24$, based on a fit to a sample of $z>4$ quasars by 
Priddey \& McMahon (2001).

\subsection{Individual Sources}

\noindent{\bf SDSS J1630$+$4012 (z=6.05)} Not detected in either the 
850$\mu m$ or 450$\mu m$ filters; also undetected at 1.2mm.\newline

\noindent {\bf SDSS J1048$+$4637 (z=6.23)} Detected at neither
850$\mu m$ nor 450$\mu m$. Based on the detection at 1.2mm
(Bertoldi et al. 2003), the SCUBA 850$\mu m$ limit is deep enough
that we should have been able to detect the source with a $4\sigma$
significance were its emission at 1.2mm to be characteristic of a
greybody at $T_D=40K$. Our non-detection thus suggests that the dust
in this object is colder than $40$K.\newline

\noindent{\bf SDSS J1148$+$5251 (z=6.43)} The most optically luminous of the
three sources, detected at both 850$\mu m$ 
and 450$\mu m$ with fluxes of $7.8(0.7)$mJy and $24.7(7.4)$mJy 
respectively -- one of very few quasars at $z>4$ where this has been achieved. 
The observed fluxes are not consistent with a 
Priddey \& McMahon (2001) single-temperature SED as can be seen  
in Figure~1. This can be seen more clearly in Figure~2 a--c, where the 
850$\mu m$/1.2mm and 450/850$\mu m$ ratios have been 
plotted as a function of 
redshift, both for the single-temperature model (locus) and for a 
selection of high-z quasars, including those observed in this sample. 
The observed flux ratios are discrepant at the $1\sigma$ level, 
however are well within the $2\sigma$ limits.\newline

\subsection{Observed Fluxes}

The three sources observed show quite different submm properties, 
in spite of their similar absolute B-band ($M_B$) magnitudes
(column [3] in Table~2),  
and so inferred blackhole/bulge masses (see column [10] in Table~2). 
This is not surprising as, to date, it has not been possible to establish a 
correlation between the optical luminosity of a quasar and the 
submm emission from its quasar host galaxy using  
samples of radio-quiet, optically loud quasars at  $z>4$ and $z\approx2$
(Isaak et al. 2002; Priddey et al. 2003). 
What is striking, however, are the detections at 
both 850$\mu m$ and 450$\mu m$ of SDSS1148$+$5251. 

It is not possible to fit the two fluxes reported here and the 1.2mm
flux with a single-temperature SED parameterised by $T_d = 40$, and
$\beta=2$ (see Figure~1: solid line). A better fit can be achieved,
with a much hotter characteristic dust temperature $T_d = 180K$, and
$\beta=0$ (Figure~1: dotted line), or using a two-component model with
a cool component characterised by $T_d=40K; \beta=0$ and a hotter
component with $T_d=140K; \beta=1.5$. However, we stress that there
are insufficient data points to constrain such a two-component model. If, in the
first instance, we assume that the underlying SED is indeed best
characterised by a single-temperature, then there are a number of
different factors that need to be explored to establish the origin
of the anomalous flux ratios, which can be broadly grouped into those
that result in an anomalously high millimetre flux, and those that may
result in systematically low submm fluxes. However, it is clear that
observations with a higher signal-to-noise are urgently needed in
order to better constrain the model fitting.  Furthermore, the high photometric precision of ALMA, of order a few percent, will be crucial for future analyses of this type.

\subsubsection{A high 1.2mm flux:}
If we assume that the observed 850$\mu m$ flux is correct, then there
are a number of observational and physical reasons why the observed
1.2mm and 450$\mu m$ fluxes might deviate from values expected for a
single-temperature SED model:

{\bf Relative calibration:} Calibration at mm and long-submm
wavelengths is relatively straightforward, particularly under periods
of high and stable atmospheric transmission. A comparison of
calibrators common to JCMT and IRAM by Lisenfeld and collaborators
suggests that the calibrator fluxes measured at 850$\mu m$ and 1.2mm
are consistent with thermal SED profiles. We have included calibration
errors of 10 -- 15\% in our plotted rms estimates. Accurate
calibration at 450$\mu m$, however, is far more difficult with small,
temporal variations in atmospheric extinction, ($\tau$), contributing significantly to the final,
overall flux. We estimate that the combined random and systematic
error in the individual flux measurements to be about 30\%. We have
included calibration errors in the error-bars plotted in Figure~1,
calculated by taking the square root of the sum of the squares of the
different error components.

{\bf Synchrotron contamination:} Synchrotron emission can contribute
significantly to the observed mm fluxes of radio-loud quasars, with
boosting from either the synchrotron tail in the quasar itself, or
from a source in its neighbourhood.  VLA observations at 43GHz by
Bertoldi et al. (2003) place a $S(3\sigma)_{43\rm GHz} < 0.33$mJy on
radio emission from the quasar at 43GHz.  This strongly suggests that
the 1.2mm emission is thermal rather than non-thermal given the
positive spectral shape. A search of the NVSS (Condon et al. 1998) and
FIRST (Becker et al. 1995) radio catalogues did not reveal
any radio sources within a 30 arcsec radius of the optical position of
the quasar. The FIRST survey has a 0.138mJy/beam rms at the quasar
position (1.4GHz), which places a $3\sigma$ upper limit to the
synchrotron contamination of the 1.2mm flux by radio sources in the
quasar field of $0.5$mJy, based on the worse case of a
flat-spectrum radio source.

{\bf Lensing:}
Gravitational lensing can boost observed flux measurements considerably
(eg. F10214$+$4724 (Broadhurst \& Lehar 1995), 
APM 08279$+$5255 (Irwin et al. 1998; Ibata et al. 1999). 
In general, however, the 
magnification is achromatic. Thus, all (sub)mm fluxes would be 
scaled by the same factor unless the physical extent of the regions 
emitting at the different wavelengths were quite different -- for example, 
if the mm flux ($\lambda_{rest}= 170\mu m$) traced a cooler and, 
most importantly, highly extended region whilst the shorter wavelength 
emission traced much more compact emission. It is not possible to 
establish this, however, without the high spatial resolution 
achievable using mm interferometers.

{\bf Spectral line contamination:} Line contamination of submm
continuum fluxes observed in local galaxies is widely recognised.
Observations by Zhu et al. (2003) have shown that the CO(3--2)
rotational line can contribute as much as 70\% to the 850$\mu m$
continuum flux. At $z\sim 6.43$ the CO(3--2) ($\lambda_{rest}= 867\mu
m$) line is redshifted to 6.46mm (46.4GHz), well outside the 1.2mm and 850$\mu
m$ filter passbands. Under good observing conditions, the 1.2mm 
filter has an effective pass-band of around $80$GHz, which at
$z\sim6.43$ includes the forbidden, rest-frame far-infrared
(FIR) transition of $C^+$ ($\lambda_{rest}$=157.74$\mu$m). Significant
flux boosting by this line is, however, unlikely: observations with
ISO have shown that the line-to-FIR ratio in local starbursts and
ULIRGs (Luhman et al. 1998, 2003) is a factor of ten or more lower
than that seen in normal galaxies (0.1--1\%: (Stacey et al. 1991)). If
we assume an equivalent filter width of the MAMBO-2 camera at 1.2mm
of around 80GHz, then the $C^+$ line peak would need to be over
500mJy to account for the excess emission at 1.2mm(1.3mJy) above
that expected for a single-temperature SED fit. This would be
equivalent to a line contribution to the FIR luminosity of about
1\%. A search by Isaak et al. (1994) for the redshifted $C^+$
emission line in the optically-luminous, radio-quiet quasar BR
B1202$-$0725 at $z=4.695$ placed a $3\sigma$ upper limit of $\sim
1.8$x$10^{-4}$ on the line contribution to the FIR luminosity
(equivalent to a $3\sigma$ upper limit to the line of $\sim 60$mJy,
and to the line-to-continuum ratio of just over 1 in a line-width of
$250\rm kms^{-1}$) in this, the most submm-bright of the $z>4$
quasars. Thus, whilst an intriguing possibility, it is unlikely that
contamination by $C^+$ is responsible for the high 1.2mm flux.
\begin{figure}

\begin{center}

\vspace{160mm} \includegraphics{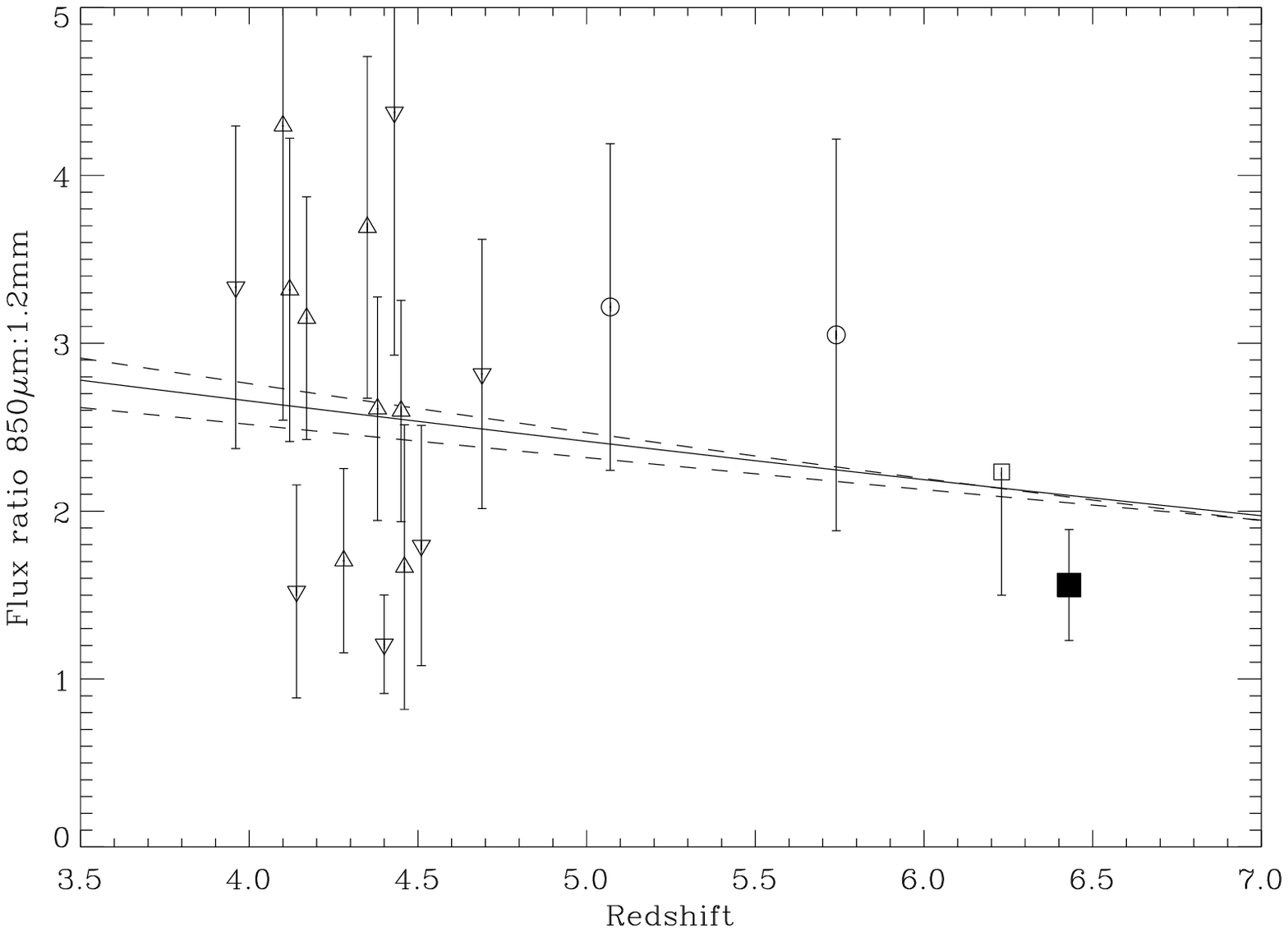}

\includegraphics{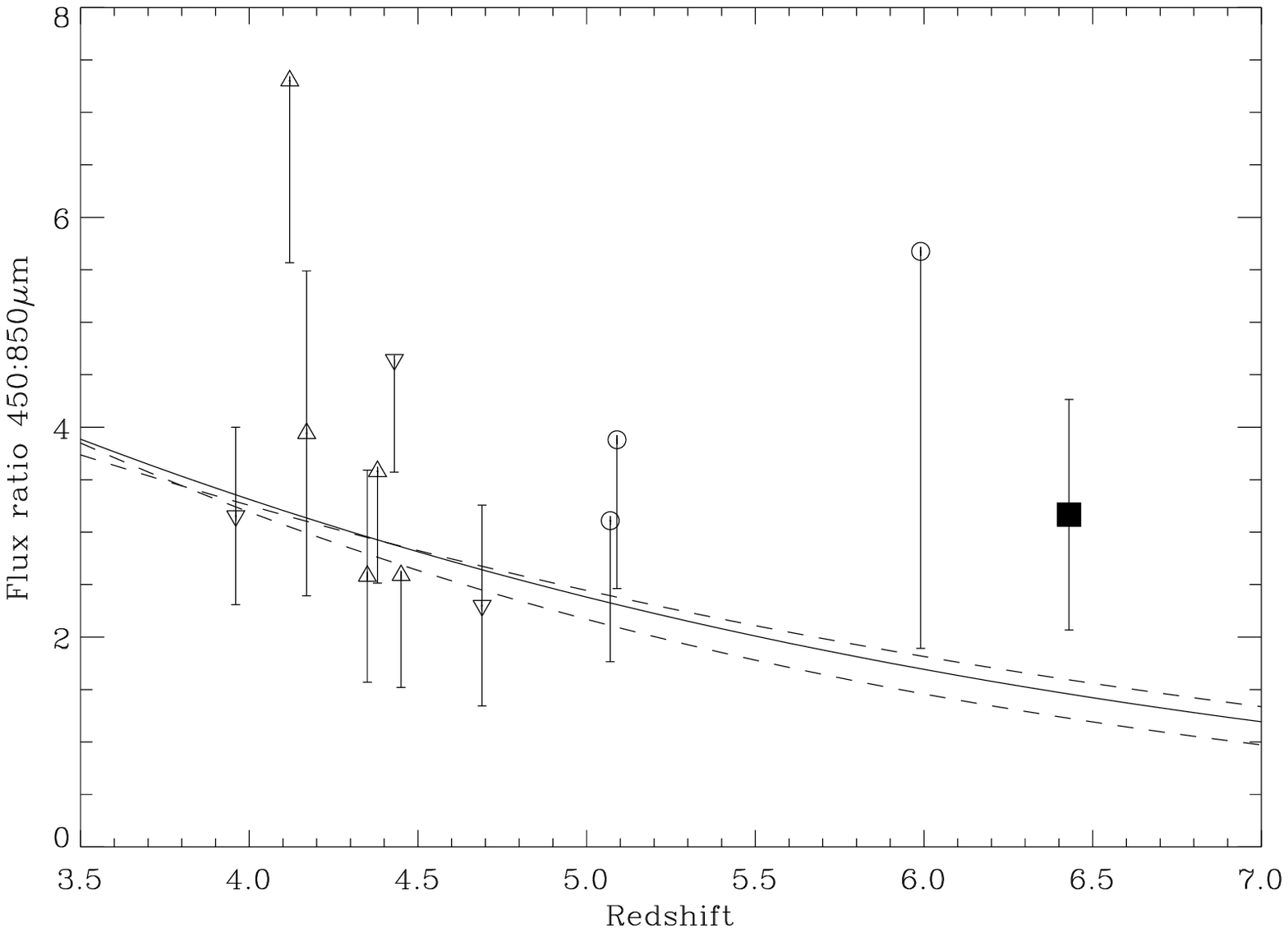}

\includegraphics{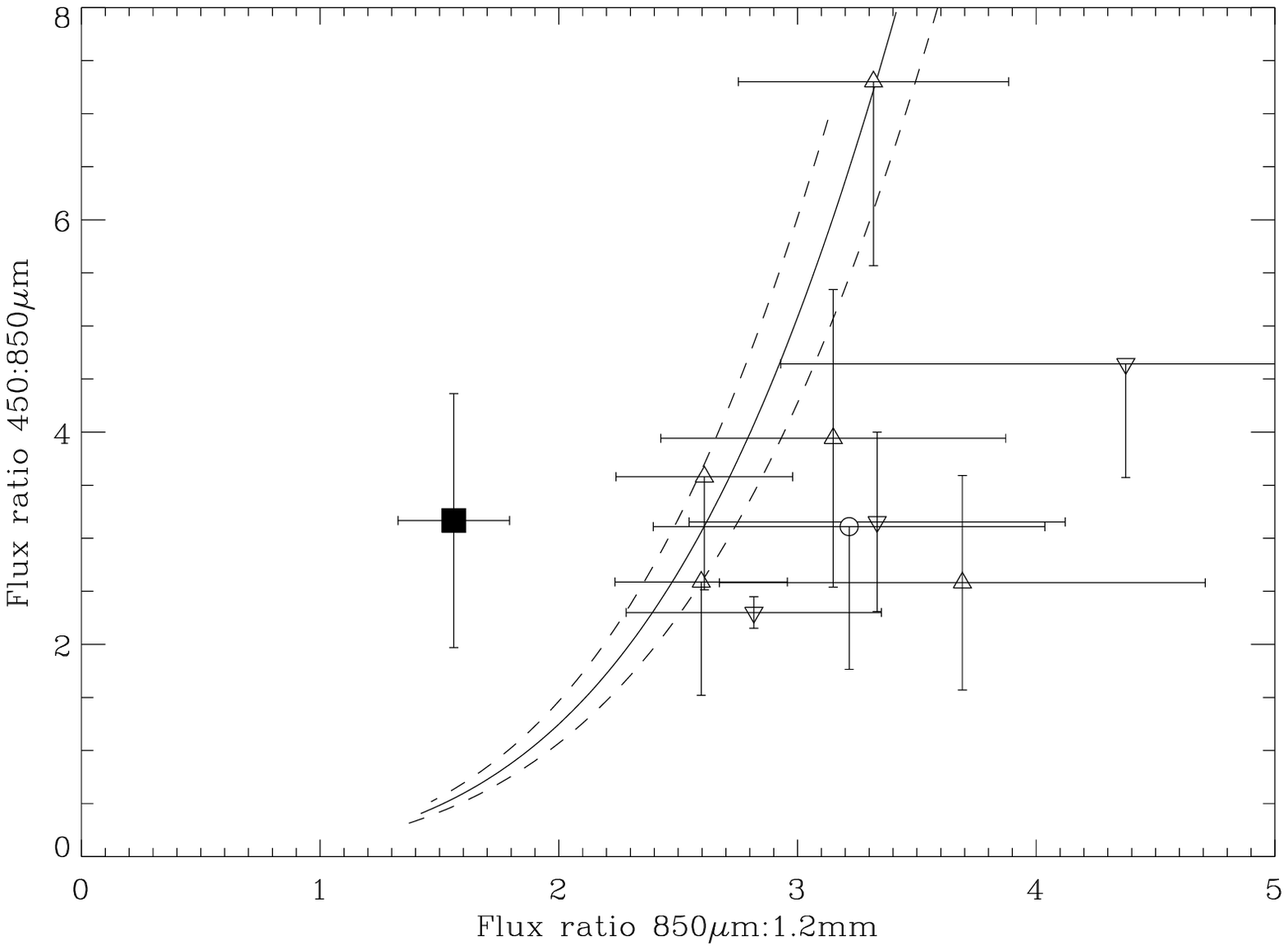}

%\special{psfile=scaled_sed.ps hoffset=110 voffset=-110 hscale=30 vscale=30 angle=0}
\end{center}

\caption{Mm and submm flux ratios: diagnostics of the underlying SEDs.
Upper panel: Ratio of $S_{850\mu m}/S_{1.2mm}$ as a function of
redshift\newline Middle panel: Ratio of $S_{450\mu m}/S_{850\mu m}$ as
a function of redshift \newline Lower panel: Ratio of $S_{450\mu
m}/S_{850\mu m}$ vs.  $S_{850\mu m}/S_{1.2mm}$ \newline In each case
the solid and dashed lines represent the locus of the ratios
calculated from the Priddey\&McMahon SED and $1\sigma$ deviations
therefrom respectively; the upright triangles represent data taken
from Isaak et al., (in prep.); the inverted triangles data taken from
Omont et al. (1996); McMahon et al. (1999); Omont et al. (2001); 
Carilli et al. (2001); Isaak et al. (2002); Bertoldi et al. (2003), 
the open circles data taken from Priddey et
al. 2003b and the filled square the detection reported in this
paper. The error bars are $1\sigma$ values.  The anomalous observed
fluxes can best be seen in the lower panel, where the ratio of the
ratios deviates by between 1 and 3$\sigma$ from the
single-temperature SED fit.  }

\end{figure}

\subsubsection{A low 850$\mu m$ flux:}
If, in contrast, we assume that the 1.2mm flux is correct, then the
observed submm fluxes at 850$\mu m$ and 450$\mu m$ are a factor of 1.4
and 0.7 lower than one would expect from a single-temperature,
$T_d=40K; \beta=2$ SED fit.  The 850$\mu m$ detection flux is at
$>10\sigma$ and as such is, at first glance, almost $4\sigma$ below
the single temperature fit derived from the 1.2mm flux. The
statistical significance of the difference however, ranges between
$0.7 < \frac{S^{est}_{850}  - S^{obs}_{850}}{\sigma^{obs}_{850}} < 6.7$ where $S^{est}_{850}$ is the $850\mu m$ flux estimated from the observed 1.2mm flux when one factors in the uncertainties in the 1.2mm flux and the error
bars in the SED fit itself. The statistical significance of the
difference between the observed and predicted 450$\mu m$ flux is much
less, as the detection itself is only at $3\sigma$. 
Interestingly, the discrepancy between the
observed 450$\mu m$ flux and that derived from the 1.2mm flux (scale
factor of 3.3 from Figures~2a--c) is smaller.\newline

Thus, the observations of SDSS1148$+$5251 suggest that the dust
emission is not well characterised by a single-temperature SED fit.
One cannot, however, attach high significance to this statement because of the relatively low value of the signal-to-noise at 450$\mu m$ in particular. Bearing
this in mind there is {\it very tentative} evidence that alludes to a
change in the properties of the high-z quasar hosts with
redshift. There is a considerable spread in the flux ratios and a
more detailed analysis of the SEDs requires not only very high
signal-to-noise data in the 450$\mu m$ filter, but also data taken at
even shorter wavelengths that match more closely the peak and
predicted turnover of the rest-frame SED.  We defer a more detailed
discussion of the SEDs of high-z quasars, as well as tentative trends
with redshift to Isaak et al. (in prep.)

\subsection{Inferred properties}
SDSS1148$+$5251 is unique for two reasons: first, on account of its
redshift; second, because it is detected not only at 1.2mm and
850$\mu$m, but, uncommonly for a high-redshift quasar, also at
450$\mu$m.  The consequence of assuming a low
$T_d$---as suggested by the 1.2mm--850$\mu$m flux ratio---is a
prohibitively large dust mass. This burdens us with explaining how so
much dust synthesis could have taken place without the formation
redshift being unacceptably high.  If, in contract, $T_d$ is high---as
suggested by the 450--850$\mu$m ratio---there is no problem
accounting for the dust mass, which is small as a result. This is,
however, at the cost of a large FIR luminosity. Indeed, if the dust
really is this hot, then the far infrared luminosity approaches the
blue luminosity of the quasar, which would suggest that reprocessed
AGN emission is not the dominant mechanism heating the dust.

Notwithstanding the uncertainty in dust temperature, Table~2 lists the
properties all the targets would have if they were ``average'' $z>4$
quasars, i.e. possessing the $T_d$=40K and $\beta$=2 fit by Priddey \&
McMahon (2001).  The cosmological parameters $\Omega_M$=0.3,
$\Omega_{\Lambda}$=0.7, $H_0$=65km/s/Mpc are assumed.
$t(\infty)-t(z)$ is the difference in lookback time between redshift
$z$ and redshift $\infty$.  As in Priddey et al. (2003b) we adopt a
dust opacity of $\kappa(\lambda)$=30cm$^2$g$^{-1}\times\lambda^{-2}$
$M_d$ is thus the mass of dust. $\dot{M}_*$(min) is the absolute
minimum sustained star formation rate needed to synthesise $M_d$
within the available time.  $M_{bh}$ and $\dot{M}_{acc}$ are the black
hole mass and accretion rate calculated from the absolute $B$
magnitude ($M_B$) assuming Eddington accretion
(eg. Isaak et al. (2002)).  We have, however, shown that observed
submm emission from SDSS1148$+$5251 is not consistent with the
Priddey et al. (2001) fit. In Figure~1 we have considered instead a
selection of alternative SEDs: the ``hot'' model ($T_d$=180K,
$\beta$=0) has ($L_{FIR}$, $M_d$)=(1.1$\times$10$^{14}$L$_{\odot}$,
0.3$\times$10$^8$M$_{\odot}$), whilst the two-component model
illustrated has ($L_{FIR}$, $M_d$)=(2.9$\times$10$^{14}$L$_{\odot}$,
3.3$\times$10$^8$M$_{\odot}$).  The ``mean'' SED, on the other hand,
gives ($L_{FIR}$, $M_d$)=(0.1$\times$10$^{14}$L$_{\odot}$,
5.3$\times$10$^8$M$_{\odot}$).

\section{Conclusions}

The rest-frame FIR spectral energy distribution 
is key to determining the thermal origin of the observed submm    
emission from high-redshift quasars. The observations presented here
suggest that the host galaxies of quasars out to redshifts of 
$z>6$ are actively undergoing star formation. Multi-wavelength observations
spanning the mm and submm are crucial to providing 
constraints on the dust mass, far-infrared luminosity and inferred star
formation rate, thus further exploring the role of star formation in 
high-redshift quasar host galaxies.  

\section*{Acknowledgments}
We would like to thank staff at JCMT and JAC for their support during 
observing, along with the referee who responded extremely 
promptly and with helpful suggestions. 

%\bibliography{/home/serpens/spxkgi/papers/bibfiles/astronomy_bibfile}

%\bibliographystyle{mn2e}

\bsp

\label{lastpage}

\end{document}